# Revisiting the size effect in software fault prediction models

Amjed Tahir
*School of Engineering and Advanced Technology, Massey University, NZ*
a.tahir@massey.ac.nz

Kwabena E. Bennin
*Department of Computer Science City University of Hong Kong, HK*
kebennin2-c@my.cityu.edu.hk

Stephen G. MacDonell
*Department of Information Science University of Otago, NZ*
stephen.macdonell@otago.ac.nz

Stephen Marsland
*School of Mathematics and Statistics Victoria University of Wellington, NZ*
stephen.marsland@vuw.ac.nz

**Abstract**

*BACKGROUND: In object oriented (OO) software systems, class size has been acknowledged as having an indirect effect on the relationship between certain artifact characteristics, captured via metrics, and fault-proneness, and therefore it is recommended to control for size when designing fault prediction models. AIM: To use robust statistical methods to assess whether there is evidence of any true effect of class size on fault prediction models. METHOD: We examine the potential mediation and moderation effects of class size on the relationships between OO metrics and number of faults. We employ regression analysis and bootstrapping- based methods to investigate the mediation and moderation effects in two widely-used datasets comprising seventeen systems. RESULTS: We find no strong evidence of a significant mediation or moderation effect of class size on the relationships between OO metrics and faults. In particular, size appears to have a more significant mediation effect on CBO and Fan-out than other metrics, although the evidence is not consistent in all examined systems. On the other hand, size does appear to have a significant moderation effect on WMC and CBO in most of the systems examined. Again, the evidence provided is not consistent across all examined systems CONCLUSION: We are unable to confirm if class size has a significant mediation or moderation effect on the relationships between OO metrics and the number of faults. We contend that class size does not fully explain the relationships between OO metrics and the number of faults, and it does not always affect the strength/magnitude of these relationships. We recommend that researchers consider the potential mediation and moderation effect of class size when building their prediction models, but this should be examined independently for each system.*

## 1. INTRODUCTION

Due to their key role in the allocation and prioritization of effort in testing and maintenance, the software engineering community has shown enduring interest in building and evaluating fault prediction models. Typically, these models employ some functions of software design and code metrics to predict faults, or fault-proneness [2, 26]. The performance of these prediction models has been the subject of extensive discussion in the literature [5, 11, 23]. In fault prediction research for classes in object-oriented (OO) systems, it is conventional to account for the confounding effect of class size. In essence, this reflects an assumption that larger classes are inherently 'more likely to be faulty', or 'more faulty', than smaller classes. This approach has become the expected convention - in submitting our research previously, the need to consider the indirect (confounding) effect of size on the relationship between software characteristics and faults was raised frequently in reviews. While this effect is intuitively appealing, to the best of our knowledge just a few prior studies have sought to empirically address this issue in any detail (e.g., [6, 9, 25]). The work of El Emam et al. [6] and Zhou et al. [25] showed that class size has a confounding effect on the relationship between most OO metrics and class fault-proneness. It was therefore suggested that empirical studies should control for size when designing fault prediction models. The recent study of Gil and Lalouche [9] suggested that code size is the only unique valid metric and therefore fault prediction models should not in fact control for size. The authors explained that the more a metric is correlated with size, the more able it is to predict bugs.

These previous works investigated the effect of size with regards to fault-proneness (i.e., a class is either faulty or it is not). We argue, however, that the conclusions presented in these studies may not apply to other fault prediction models - especially those that seek to build continuous models (i.e., predicting the *number* of faults) – due to the statistical methods used. In particular, there are problems with the assumption of normality of data distributions, and also regarding the power and Type I error control of the tests employed in both studies. As such their conclusions may not hold.

The need to adopt more robust statistical methods in empirical software engineering research has been emphasized recently by Kitchenham et al. [17], as has the (consequent) need to reassess evidence from prior empirical results. The goal of this work is to revisit the effect of size on the association between OO metrics and the number of faults in a class using robust statistical methods. The focus of this work is on continuous fault prediction models; that is, models that predict the number of faults in a class or a file. Specifically, we conduct detailed *mediation* and *moderation* statistical analyses to investigate the possible effect of class size on the relationships between seven OO metrics and the number of faults. To the best of our knowledge, our work is the first to examine the effect of size in continuous fault prediction models. Based on the analysis conducted in this study, we show that: *1) there is limited evidence of a mediation effect of size on the association between some OO metrics and the number of faults*. In contrast to prior studies on fault-proneness, the mediation effect is not evident across all systems examined. Alternatively, we found that *2) there is more evidence that size has a significant moderation effect, especially for CBO and Fan-out metrics, as it impacts the strength/magnitude of the relationship between OO metrics and faults*. However, this moderation effect displays a lack of consistency as it does not appear to affect all individual systems.

Our study complements previous studies in that it is the first to address the issue of the effect of size in continuous fault prediction models. We also use contemporary robust statistical procedures to accurately estimate the true effect of size in these prediction models. Instead of only estimating the indirect or mediation effect, we provide evidence by analysing both the *mediation* and *moderation* effect of size in fault prediction models.

## 2. RELATED WORK

El Emam et al. [6] empirically investigated the relationship between OO metrics and class fault-proneness. In conducting their research the authors built two logistic regression models for each OO metric considered, one included LOC (as the size measure) while the other did not. They further calculated two odds ratios for the two models, which respectively measured the magnitudes of the association between the OO metric and fault-proneness with and without controlling for size. They then compared the values of the two odds ratios âĂŞ if there was a large percentage difference in the two values then class size was considered to have a confounding effect. The results of the study indicated that the associations between most of the metrics investigated (namely, Weighted Methods per Class (WMC), Response for Class (RFC) and Coupling Between Objects (CBO)) and fault-proneness were significant before controlling for size but were not so after controlling for size. The authors thus recommended that fault prediction models should always control for size. This approach mirrors the Baron and Kenny casual-steps (joint significance) test [12]. While simple and widely used, the casual-steps test is known to have less power than other available mediation analysis approaches [8, 18]. Additionally, this approach is not based on a quantification of the intervening indirect effect, but examines a series of links in a casual chain between variables (e.g., $X \to M \to Y$). A similar approach was used by Gil and Lalouche [9], which also followed a casual-steps approach to determine the effect of size on a number of OO metrics. However, the study found that, when controlling for size, metrics may lose their predictive power. Evanco [7] also argued that introducing class size as an additional independent variable in a fault prediction model can result in misspecified models that lack internal consistency.

Zhou et al. [25] also addressed the effect of class size on the association between OO metrics and fault-proneness. The authors employed the Sobel test, which examines the significance of the mediation effect by calculating the standard error of $XY$, evaluating the null hypothesis that the true effect of a mediating variable is zero. Their results showed that size has a mediation effect on the relationship between OO metrics and fault-proneness. Positively, the Sobel test is known to perform better (in terms of power) than the casual-steps approach [8]; however, there are a number of issues with this test, particularly its weakness due to the assumption of normality of sampling. In most cases the sampling distribution of $X \to Y$ tends to be asymmetric, with high skewness and kurtosis [4, 18] (In our analysis of one of the two datasets used in [25] (originally generated in [5]), we found that the data were not normally distributed (see Section 3). In addition, the Sobel test has also been shown to score low in power and Type I error control [18].

Note that all previous studies have investigated the size effect from only one perspective, the indirect effect of size; that is, where size is a third variable and that the relationship between the metrics and fault-proneness is transmitted through this variable. As such, size is in this case a *mediator* variable, and such a variable is analysed using a mediation analysis techniques (i.e., studying the indirect effect).

These known issues with both techniques (i.e., the casual-steps and Sobel tests), combined with a level of conclusion instability across different studies, led us to seek alternative approaches to investigate the true indirect effect of size. In this work we assess whether class size has a mediation effect on the relationships between OO metrics and faults (in continuous models) using robust statistical analysis. In addition, we assess another size effect phenomenon known as the *moderation* effect - we study whether class size moderates (i.e., affects the size or strength of) the relationship between OO metrics and the number of faults. We are not aware of any other study that has investigated the moderation effect of class size in fault prediction models. As such our results should inform the design of future models that may - or may not - employ class size as one of the predictors of faults.

## 3. RESEARCH SETTING AND EMPIRICAL ANALYSIS

### 3.1 Estimating the indirect effect of variables

Methods to detect the indirect effect of confounding variables have been discussed widely in other disciplines such as psychology and epidemiology.

Given an independent variable $X$ and a dependent variable $Y$, the effect of $X$ on $Y$ may be transmitted through a third intervening (mediating) variable $M$. That is, $X$ affects $Y$ because $X$ affects $M$ and $M$ in turns affects $Y$ (i.e., $X \to Y$ is the result of the indirect relationship $X \to M \to Y$). Fig. 1 depicts a path diagram showing a direct relationship between an independent variable (IV, an OO metric) and a dependent variable (DV, number of faults). Fig. 2 depicts the possible indirect relationship between an OO metric and the number of faults through a mediator: class size. If a mediator is controlled (i.e., held constant) and that mediator was the reason for the relationship between the independent and dependent variables (i.e., there is complete mediation), then the direct relationship between the independent and dependent variables becomes zero.

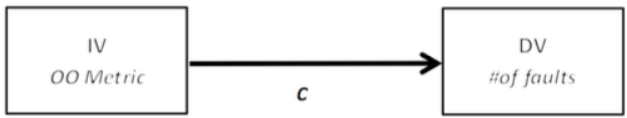

Figure 1: Path diagram illustrating a simple model for the relationship between OO metrics and faults

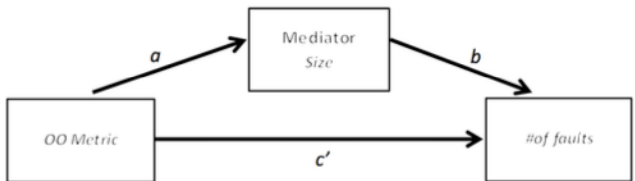

Figure 2: Path diagram illustrating the mediation effect of size on the relationships between OO metrics and faults

On the other hand, a *moderator* is a variable that impacts the power of the relationship between the independent and dependent variables. It can affect the sign or the strength/magnitude of the relationship between those two variables. Fig. 3 shows the potential moderation effect of size on the relationship between an OO metric and the number of faults. The relationship between $X$ (independent variable) and $Y$ (dependent variable) is said to be moderated if its size or direction depends on a third variable $M$ [13]. In other words, a moderator is a third variable that affects the zero-ordered correlation between the independent and dependent variables. Unlike mediation, which affects to the extent that it *accounts for* the relationship between the independent and dependent variables, a moderation variable does not necessarily explain the relationship between variables but rather it explains *the strength of* this relationship [1]. A moderator effect is indicated by the interaction of $X$ and $M$ in explaining $Y$.

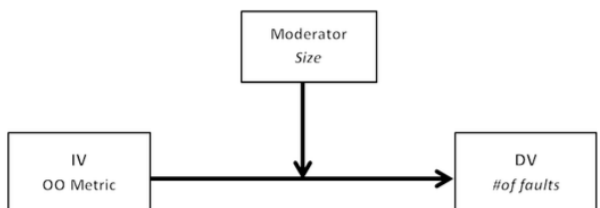

Figure 3: Path diagram illustrating the moderation effect of size on the relationships between OO metrics and faults

Note that all previous studies have investigated the size effect from only one perspective, the indirect effect of size; that is, where size is a third variable and that the relationship between the metrics and fault-proneness is transmitted through this variable. As such, size is in this case a *mediator* variable, and such a variable is analysed using a mediation analysis techniques (i.e., studying the indirect effect).

### 3.2. Methods to detect mediators and moderators

A number of candidate approaches can be employed to determine the true effect of a mediating variable on the relationship between two other variables, and MacKinnon et al. [18] provide a detailed comparison of several well-known methods. Of those methods, there is a strong and quite recent recommendation in the literature to use bootstrapping-based mediation tests over other tests such as the casual-steps and Sobel tests [12, 18].

Bootstrapping-based methods for estimating indirect effect have been discussed in the literature for some time (e.g., [4]), but the methods have received more attention in recent years [12, 19, 20]. Bootstrapping is a nonparametric data re-sampling procedure that generates an empirical representation of the original sampling distribution in order to make inferences, rather than making assumptions about the original population. Bootstrapping repeatedly resamples (with replacement from the original data points) the obtained sample of size $nk$ times during analysis, as a means of mimicking the original sampling process. The indirect effect is then computed from each of these samples. This bootstrapping approach tends to have better power and Type I error control than other approaches [12, 18]. In fact, bootstrapping has been described as the best available method overall for assessing mediating variables [19], preferred over classical methods such as Sobel test. In this work, the bootstrapping-based method is used to test for different effects multiple times in both moderation and mediation analysis, via regression-based approaches [13]. In our analysis, mediation analysis is estimated through three types of effect: the *direct*, *indirect* and *total* effect. We estimate all three effects using the following rules (as illustrated in Figs 1 and 2):

- Total effect of $X$ on $Y = c$
- Direct effect of $X$ on $Y = c\,'$
- Indirect effect of $X$ on $Y$ through $M = ab$

In moderation analysis, we study whether the effect of the increase/decrease in the independent variable is not dependent on the moderator. We measure the change in the R-square value due to the interaction of the moderator variable. In case that such an interaction is significant, we then probe the interaction of moderators using two techniques: 1) simple slope analysis and 2) the Johnson-Neyman method. Simple slope analysis can visually portray how another variable moderates the association between the independent and dependent variable. The conditional effect of a moderator variable $M$ on the relationship between an independent variable $X$ on a dependent variable $Y$ is estimated as follow:

- Conditional effect of $X$ on $Y = c\,' + bM$

While both effects (mediation and moderation) are estimated differently, it is still possible that a variable can have both a mediation and moderation effect [1] (i.e., a mediator $M$ interacts with the relationship between the independent $X$ and the dependent variables $Y$ ). Such variables are known as *moderated mediation* variables [21]. This work aims to estimate both mediation and moderation effects of size, assuming that size might have both effects on the relationship between OO metrics and the number of faults in a class. Note that the previous studies in this space (e.g., [6] and [25]) have investigated the *mediation* effect of size in fault prediction models. This is what has been refereed to as the "confounding effect". While it is important to confirm whether size has a mediation effect in continuous fault prediction models, we also believe that it is equally important to investigate the potential moderation effect of size (that is, the impact of size on the strength of the relationship between OO metrics and the number of faults).

### 3.3 Conducting mediation and moderation analysis

As noted above, in this study mediation analysis is conducted using a bootstrapping-based procedure [12, 14]. After conducting a series of regression analyses between all variables, bootstrapping is used to calculate the associated confidence interval. Let us say we take 1000 bootstrap samples, the point estimates of variables $X$ and $Y$ are the means computed over those samples and the estimated standard error is the standard deviation of the 1000 $XY$ estimates. We then compute a confidence interval using the Monte Carlo method (advantages of the Monte Carlo method over other similar methods are discussed in [22]). To derive a 95% confidence interval, the elements of the vector of 1000 estimates of $XY$ are sorted into ascending order. The Lower Limit (LL) of the confidence interval is the 25th score in this sorted distribution, and the Upper Limit (UL) is the 976th score in the distribution. If zero does not fall between the resulting confidence interval values of LL and UL (i.e., both values are either positive or negative) then it is concluded that the indirect effect is not zero with 95% confidence [20]. That is, the possibility of a mediation effect is significant.

In applying bootstrapping mediation to our analysis of the relationships between several OO metrics, LOC and the number of faults, we employ 5000 bootstrap re-samples and set the confidence intervals to 95%, as per the recommendation in [20]. These procedures were carried out using the PROCESS SPSS macro[1]. PROCESS provides a mechanism to conduct both mediation and moderation analysis using ordinary least squares (OLS) regression analysis. The mathematical definition of the methods used is outside the scope of this work, however readers may refer to [13] for more information. We use $R$ for all other statistical analysis.

We apply our moderation and mediation analysis procedures to five OO metrics, LOC and faults. For each system in our datasets, the total number of tests conducted is $2 \times 7 = 14$ tests for both moderation and mediation analysis. For reproducibility and replication, we provide all scripts in the supplementary material online [24]. For mediation analysis, PROCESS produces a series of regression models. In the first model the dependent variable (number of faults) is explained directly by an independent variable, i.e., a single OO metric (step 1). This is called the *direct effect* (see Fig. 1). Then the mediator (LOC) is explained using the independent variable (the same single OO metric) (step 2). Another regression model is then built to explain variance in the dependent variable (number of faults) using both the independent variable and the mediator (step 3) - this is the path $ab$ in Fig. 2. The *indirect effect* is then estimated. Note that we estimate the indirect effect (at step 3) only if the first two steps are found to be significant (i.e., the *total* and *direct* effects are both significant). If all three steps are found to be significant, then we establish that data are consistent with the hypothesis that variable $M$ mediates the $X{\rightarrow}Y$ relationship. On the other hand, if any of the first two steps were not found to be significant, then we conclude that the mediation effect of $M$ is unlikely.

Subsequently, we conduct moderation analysis by building a regression model to model-fit the outcome (number of faults) using the independent variable (an OO metric), the moderator and an interaction term which tests whether the relationship between the OO metric and outcome changes based on the moderator. The interaction term is computed as the product of the independent variable and the moderator. By using the PROCESS tool, we conduct moderation analysis of each OO metric in explaining the number of faults. For a significant moderation, we analyze the regression table produced by PROCESS. If the interaction term (denoted as $int_1$) has a $p < .05$ and the bootstrapping-based confidence interval values of LL and UL do not include zero, the moderation effect is then declared as significant with 95% confidence.

### 3.4 Datasets

To examine the possible size effect we selected two publicly available open source datasets from D'Ambros et al. [5] and Jureczko and Madeyski [16]. General information about the systems in both datasets (combined) is shown in Table 1. Both datasets contain all relevant metrics and faults data that are needed for this study. They are both publicly available and have been widely used in defect prediction studies (e.g., [3, 15]). The D'Ambros et al. [5] datasets include metrics data collected from five large open source projects from two well-established ecosystems: Eclipse (JDT Core, PDE UI, Mylyn and Equinox) and Apache (Lucene). The study of Zhou et al. [25] used the same dataset, which offers us an opportunity to com- pare our results. For simplicity, we refer to this dataset as DAMB in the rest of this paper. The Jureczko and Madeyski[16] dataset contains data from 15 open source systems and 6 proprietary software projects. However, we included only the open source systems of this dataset in our analysis so that we could compare the results with those obtained from the DAMB other dataset (also open-source) without concern over a source effect. In addition, the sample size for those 6 proprietary software projects is very limited. Finally we also excluded systems comprising small numbers of classes (fewer than 60) as we consider them too small for our analysis. The total number of systems in this dataset that we included in our study is 12. This includes 11 well-known Apache projects (including Tomcat, Ant, Log4j and Ivy) and jEdit (a popular coding editor). We refer to this dataset as JURE.

---
[1] http://www.processmacro.org

Table 1: An overview of the datasets used

| Dataset | #projects | NOC | LOC | #faults | #faulty classes | %faulty classes |
|---|---|---|---|---|---|---|
| DAMB | 5 | 5238 | 639,826 | 1,740 | 843 | %16 |
| JURE | 12 | 5736 | 1,744,794 | 4,741 | 2,329 | %40 |
| Total | 17 | 10,974 | 2,384,620 | 6,481 | 3,172 | %29 |

## 4. RESULTS AND DISCUSSION

We studied the effect of class size (measured using LOC) on the relationship between each of the following OO metrics and the number of faults: RFC, WMC, CBO, DIT, LCOM, Fan-in and Fan-out. Our research questions are:

- **RQ1:** does class size have a *mediation* effect on the relationships between OO metrics and the number of faults?

- **RQ2:** does class size have a *moderation* effect on the relationships between OO metrics and the number of faults?

We test the following two Null Hypotheses:

- $H_0$: class size has no *mediation* effect on the relationship between OO metrics and the number of faults.

- $H_1$: class size has no *moderation* effect on the relationship between OO metrics and the number of faults.

We first examined the normality of the various data distributions for the number of faults using the Shapiro-Wilk test. For the combined datasets DAMB and JURE we test the null hypothesis that the samples come from a normal distribution. The results of this test lead us to reject the null hypothesis that the samples come from a normal distribution (DAMB: W = 0.334, $\rho < 0.01$; JURE: W = 0.524, $\rho < 0.01$). We also examined the normality of distribution (for the number of faults) in all individual systems, and the results were similar. This is also visibly evident in the Kernel Density plots presented in Fig. 4. The plots visually confirm that both datasets do not have a normal shape; most faults are concentrated in certain classes or files. The classes with very prominent modes have large class size. The lower mode classes have small size.

We followed this with the three-steps approach (Section 3.2) to investigate the effect of size. We first examine the associations between OO metrics and the number of faults. Given that the data are not normally distributed, we use the nonparametric Spearman's rho ($\rho$) correlation coefficient, and correlation strength is interpreted using Cohen's classification (low correlation when $0 < \rho \leq .29$, medium when $.30 \leq \rho \leq .49$ and high when $\rho \geq .50$). We also examine the correlation between the size metric (measured in LOC), OO metrics and the number of faults.

Table 2: List of metrics used

| Metric | Full name | Metrics | Full name |
|---|---|---|---|
| LOC | Line of Code | CBO | Coupling between Objects |
| RFC | Response for Class | DIT | Depth of Inheritance Tree |
| WMC | Weighted Method Count | LCOM | Lack of Cohesion in Methods |
| Fan-in | #other classes that reference the class | Fan-out | #other classes referenced by the class |

### 4.1 The correlation between size and OO metrics

We looked into the correlations between size, OO metrics and the number of faults across the aggregated systems for each of the DAMB and JURE datasets, shown in Fig. 5 and Fig. 6 respectively. The results show that class size is significantly correlated with several of our OO metrics. LOC is also significantly correlated with the number of faults (DAMB $\rho$= .39, p= .00; JURE $\rho$= .25, p= .00). It was also found that all of the OO metrics except DIT are strongly correlated with the number of faults[2].

We then used regression to fit models that would enable us to assess the mediation and moderation effects of size on the number of faults. In this analysis regression is used to explain and estimate the strength of the effect of independent variables on the dependent variable, rather than to predict outcomes. We used each OO metric as an independent variable and the number of faults as the dependent variable. Given the nature of the each system and the way the data was collected, we provide results and analysis of the each system separately.

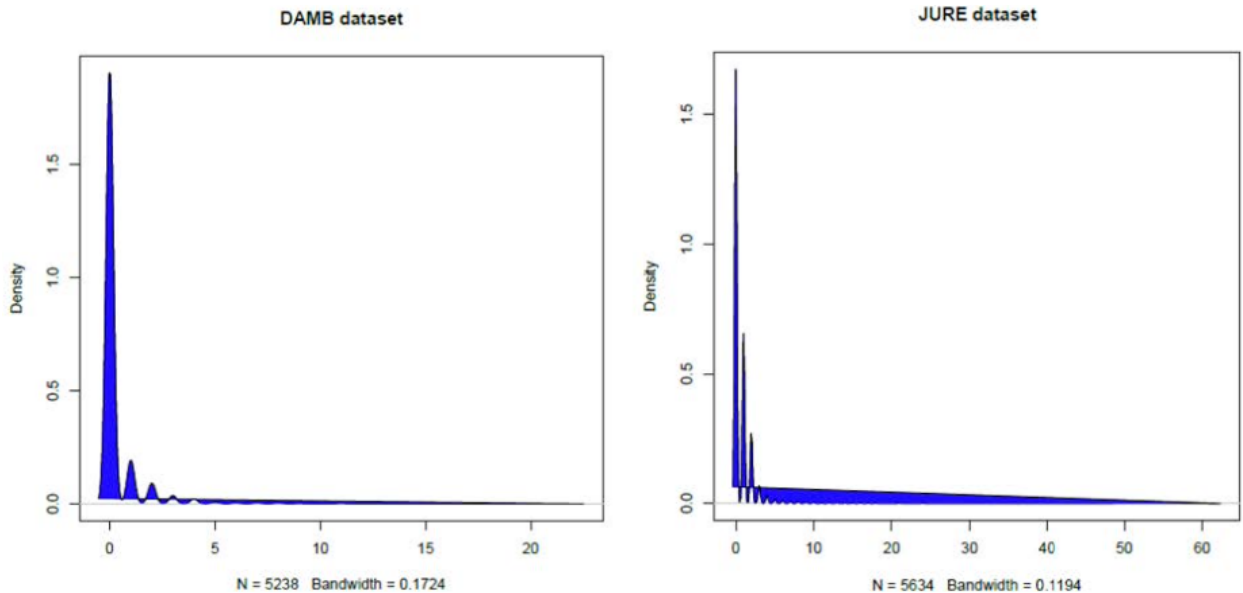

Figure 4: Kernel Density plots of faults in both datasets

---

[2] all variables are statistically significant at an alpha level 0.05.

From Tables 3 and 4 (and other tables in [24]), we observe that all of the OO metric models were statistically significant, with $P < .05$. However, for an OO metric to potentially explain variance in the outcome (number of faults), the 95% CI for each metric should not include zero. As such, all OO metrics with the exception of DIT were found to have some explanatory capability in terms of the number of faults. Fan-out in particular has significant explanatory power across many systems, with $R^2$ values of 0.46 for Xerces and 0.42 for Lucene (2.4). RFC also exhibits explanatory power, with $R^2$ values of 0.33 for Lucene (2.4) and 0.48 for Xerces.

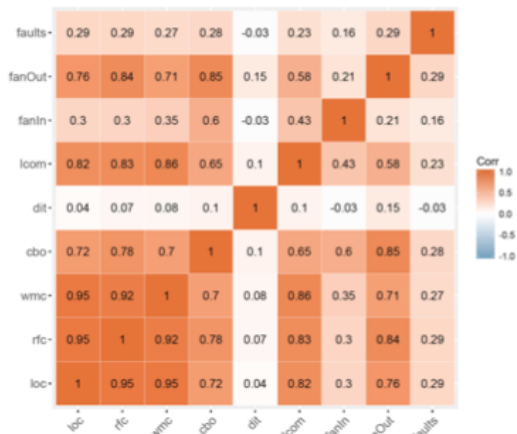

Figure 5: Spearman's correlation matrix between size, faults and OO metrics in DAMB dataset

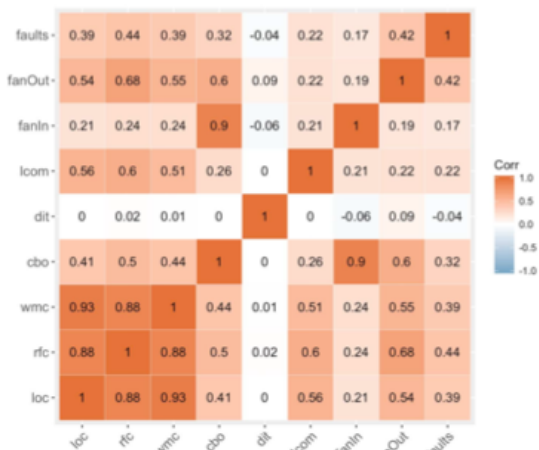

Figure 6: Spearman's correlation matrix between size, faults and OO metrics in JURE dataset

### 4.2 The mediation effect of class size

Having confirmed that the OO metrics, size metric and number of faults are all significantly correlated, and that some of these OO metrics can explain (to varying degrees of accuracy) variability in the number of faults, we turn our attention to investigate whether size plays a mediating role in the relationships between the OO metrics and the number of faults. As we explained in Section 3, the mediation effect of variables is estimated based on three consequential steps - first by estimating the total effect c (Fig. 1), then the *direct effect c',* and finally the *indirect effect* ab (Fig. 2). We explain our results below following these three steps. Note that we did not conduct any mediation or moderation analysis for DIT. While DIT is significantly correlated with the number of faults (with a rather weak negative correlation), it does not significantly explain the number of faults (Lucene (2.4): $\rho = .07$, $R^2 = .012$; Xerces: $\rho = .22$, $R^2 = .003$) or LOC (Lucene (2.4): $\rho = .41$, $R^2 = .011$; Xerces: $\rho = .20$, $R^2 = .003$). This violates the first and second steps of our three-step procedure. Therefore DIT was eliminated from any further analysis. Due to space constraints, we provide here our results from only two systems (one from each dataset), and we include results from other individual systems in the supplementary material [24].

Full results of the mediation analysis conducted on Apache Lucene (2.4) from the DAMB dataset are shown in Table 3. The *total effect* (c) regression model for RFC explaining the number of faults, ignoring the mediator, was significant ($\rho < .01$, $B = .0073$, $R^2 = .326$, $t(277) = 11.59$). The regression of RFC to explain the mediator, LOC, was also significant ($\rho < .01$, $B = 1.92$, $R^2 = .935$, $t(277) = 63.14$). The direct effect (c') of RFC on faults - by introducing the mediator (LOC) as a further explanatory factor in the model - is also significant ($\rho < .01$, $B = .004$, $R^2 = .43$, $t(276) = -3.68$). The indirect effect of RFC on faults intervals does not includes zero ($B = .0158$, BootLLCI= .0071, BootULCI= .0263); i.e., zero does not fall between the bootstrapping LLCI and ULCI values. Therefore, it is inferred that size fully mediates the relationship between RFC and number of faults. The same inference applies to the CBO, LCOM, Fan-in and Fan-out metrics. For example, for Fan-out, the total effect ($B = .102$, $R^2 = .424$) and direct effect ($B = .063$, $R^2 = .451$) are found to be significant (with $\rho < .01$), and the indirect effect intervals does not include zero (B= .040, BootLLCI= .0130, BootULCI= .0768), suggesting that class size has a significant mediation effect on the relationship between Fan-out and the number of faults. Therefore, it is concluded that class size fully mediates the relationship between the number of faults and the following metrics: RFC, CBO, LCOM, Fan-in and Fan-out in this system. On the other hand, while the total effect (c) of WMC on the number of faults was found to be significant ($\rho < .01$, $B = .015$, $R^2 = .367$), the direct effect (c') was not significant ($\rho = .33$). Failing in the second step suggests that the mediation effect of class size is unlikely (and further the indirect effect shows that zero falls between the LL and UL values).

The results of the mediation analysis from Apache Xerces (JURE dataset) are shown in Table 4. The regression model that uses RFC to explain variance in the number of faults while ignoring the mediator, i.e., the *total effect*, was significant: ($\rho < .01$, $B = .141$, $R^2 = .482$, $t(586) = 23.369$). The regression of RFC to explain the mediator, LOC, was also significant ($\rho < .01$, $B = 18.540$, $R^2 = .645$, $t(586) = 32.622$). The *direct effect* (c') of RFC on faults - by introducing the mediator (LOC) as a further explanatory factor in the model - is also significant ($\rho < .01$, $B = .145$, $R^2 = .483$, $t(585) = 14.341$). However, the *indirect effect* of RFC on the number of faults was not significant as zero falls between the interval values ($B = -.0043$, BootLLCI= -.0362, BootULCI= .0404), which indicates that (and unlike the results in Lucene) *size does not mediate* the relationship between RFC and number of faults.

Table 3: Mediation analysis results for Eclipse Lucene (2.4) (DAMB dataset)

| Metrics | B (Total Effect) | $R^2$ | F | P | SE(B) | 95% CI LL | 95% CI UL | Direct Effect ($c'$) | Indirect Effect (ab) | BootCILL[3] | BootCIUL | Sig. mediation effect |
|---|---|---|---|---|---|---|---|---|---|---|---|---|
| RFC | .0073 | .3264 | 134.216 | .00 | .0006 | .0061 | .0086 | -.0084 | .0158 | .0071 | .0263 | ✓ |
| WMC | .0148 | .3671 | 160.673 | .00 | .0012 | .0125 | .0171 | -.0046 | .0194 | -.0137 | .0446 | X |
| CBO | .0954 | .4797 | 255.410 | .00 | .0060 | .0837 | .1072 | .0697 | .0257 | .0085 | .0485 | ✓ |
| LCOM | .0026 | .3926 | 179.041 | .00 | .0002 | .0022 | .0029 | .0013 | .0012 | .0000 | .0027 | ✓ |
| Fan-in | .1315 | .1776 | 59.801 | .00 | .0170 | .0980 | .1650 | .0894 | .0422 | .0154 | .0890 | ✓ |
| Fan-out | .1025 | .4242 | 204.098 | .00 | .0072 | .0884 | .1167 | .0630 | .0395 | .0130 | .0768 | ✓ |

Table 4: Mediation analysis results for Apache Xerces (JURE dataset)

| Metrics | B (Total Effect) | $R^2$ | F | P | SE(B) | 95% CI LL | 95% CI UL | Direct Effect ($c'$) | Indirect Effect (ab) | BootCILL | BootCIUL | Sig. mediation effect |
|---|---|---|---|---|---|---|---|---|---|---|---|---|
| RFC | .1407 | .4824 | 546.109 | .00 | .0060 | .1288 | .1525 | .1449 | -.0043 | -.0362 | .0404 | X |
| WMC | .2118 | .2005 | 146.967 | .00 | .0175 | .1775 | .2461 | .0794 | .1323 | .0670 | .2328 | ✓ |
| CBO | .2889 | .2543 | 199.878 | .00 | .0204 | .2488 | .3291 | .1786 | .1103 | .0624 | .1810 | ✓ |
| LCOM | .0083 | .1145 | 75.742 | .00 | .0010 | .0064 | .0101 | .0021 | .0062 | .0029 | .0129 | ✓ |
| Fan-in | .1560 | .0459 | 28.2062 | .00 | .0294 | .0983 | .2136 | .0824 | .0736 | .0378 | .1371 | ✓ |
| Fan-out | .6546 | .4615 | 502.193 | .00 | .0292 | .5972 | .7119 | .5370 | .1176 | .0261 | .2374 | ✓ |

Table 5: Potential mediation effect of class size in all individual systems from both DAMB and JURE datasets

| | DAMB dataset | | | | | JURE dataset | | | | | | | | | | |
|---|---|---|---|---|---|---|---|---|---|---|---|---|---|---|---|---|
| | JDT | Mylyn | PDE | Equinox | Lucene (2.4) | Ant | Ivy | jEdit | Log4j | Lucene (2.2) | POI | Prop | Synapse | Tomcat | Velocity | Xalan | Xerces |
| RFC | X | X | X | ✓ | ✓ | ✓ | X | X | X | ✓ | X | X | X | ✓ | ✓ | ✓ | X |
| WMC | X | X | X | X | X | ✓ | X | X | X | ✓ | X | X | X | ✓ | ✓ | ✓ |
| CBO | ✓ | ✓ | ✓ | ✓ | ✓ | ✓ | X | X | X | ✓ | ✓ | X | ✓ | ✓ | X | ✓ | ✓ |
| LCOM | X | X | X | X | ✓ | ✓ | ✓ | X | X | ✓ | X | X | X | ✓ | ✓ | ✓ |
| Fan-in | ✓ | ✓ | ✓ | ✓ | ✓ | ✓ | X | X | X | X | X | X | ✓ | X | X | ✓ |
| Fan-out | ✓ | ✓ | X | ✓ | ✓ | ✓ | ✓ | X | X | X | ✓ | X | ✓ | N/A | ✓ | ✓ | ✓ |

However, the other metrics in Xerces show that class size indeed mediates their relationships with the number of faults. For example, Fan-out is related to the number of faults ($\rho < .01$, $B = .655$, $R^2 = .462$, $t(586)=22.409$) and the mediator LOC ($\rho < .01$, $B = .537$, $R^2 = .482$, $t(586)= 214.336$); however, the indirect effect of LOC on the relationship between Fan-out and the number of faults is not zero with 95% confidence ($B = .118$, BootLLCI= .026, BootULCI= .237), and therefore it is inferred that class size fully mediates the relationship between Fan-out and faults. The same conclusion is shared with all of the other metrics (i.e., WMC (indirect effect: $B = .132$, BootLLCI=.067, BootULCI=.233), CBO ($B = .110$, BootLLCI=.062, BootULCI=.181), LCOM ($B = .006$, BootLLCI=.003, BootULCI=.013) and Fan-in ($B = .074$, BootLLCI=.038, BootULCI=.137)) where there is evidence of significant total, direct and indirect effects.

We also analyzed the mediation effect of size in all individual systems in both datasets. Due to space constraints, we provide only summary results of this analysis in Table 5, which shows only whether size has a mediation effect for each single metric. We provide detailed tables for each system in the supplementary material [24]. As shown in Table 5, there are mixed results regarding the true mediation effect of class size across all examined systems. Considering the results for the DAMB dataset, we observed that size appears to have a mediation effect on the relationship between CBO and Fan-in metrics, and the number of faults in all 5 systems. Also, mediation effect of size on the relationship between Fan-out and faults appeared in 4 of the 5 examined systems (all systems except PDE). LCOM and RFC metrics show more mixed results, as the indirect effect appears to be significant in only 1 (Lucene) and 2 (Equinox and Lucene) systems, respectively. However, the mediation effect of size on WMC was not significant in any examined system in this dataset. When comparing the results we obtained by analyzing the DAMB dataset with those found prior [25] (using the same dataset), we found similar evidence of a mediation effect of size for the following metrics: CBO, Fan-in and Fan-out. However, unlike the previous analysis [25], we could not confirm if the effect of size is significant for the RFC, WMC and LCOM metrics.

The results of our analysis of the individual systems in the JURE dataset show similarly mixed outcomes. We observed that Apache Ant shows a mediation effect of size for all metrics, whereas jEdit, Log4J and Prop shows no mediation effect for any of these metrics. The mediation effect of size on the relationship between CBO, LCOM and Fan-out metrics, and the number of faults appeared in 7 of the 12 systems. For CBO, the mediation effect of size was not evident in Ivy, jEdit, Log4j, Prop and Velocity. For LCOM, the mediation effect of size was not evident in

jEdit, Log4J, Prop, Synapse and Tomcat. Fan-out shows similar results, with no evidence of a mediation effect of size in the following systems: jEdit, Log4j, Lucene (2.2), Prop and Tomcat. Unlike Fan-in results from the DAMB dataset, the mediation effect of size was significant only in 3 of the 12 systems (i.e., Ant, Tomcat and Xerces).

In general, it is observed that the evidence regarding the mediation effect of size is inconsistent and does not follow a pattern across all systems. For all metrics, class size is not shown to have a significant mediation effect across all systems in both datasets. A few metrics (i.e., CBO and Fan-out) show a more significant mediation effect of size than others, where a significant indirect effect was found in 12 and 11 systems, respectively. However, we still consider these results inconclusive in determining if the indirect effect is truly and always significant.

In returning to our research hypotheses (H0), based on the evidence resulting from our analysis of the systems in the two datasets individually, we cannot reject the null hypothesis that class size has no mediation effect on the relationship between the number of faults and the following OO metrics: RFC, WMC, LCOM and Fan-in. For CBO and Fan-out, while the mediation effect appears in the majority of systems, the evidence of this effect is not consistent across both datasets. Therefore, we are also unable to confirm if the mediation effect will always be present. We consider the mediation effect of size to depend largely on the systems examined. For some systems, the mediation effect of size affected all metrics, whereas for other systems there was no significant mediation effect. This may be due to the size and complexity of each specific system, or the dynamic of each development team.

**4.3 The moderation effect of class size**
To investigate the moderation effect of class size on the number of faults, we empirically assessed the impact of all six OO metrics on the number of faults under the influence of class size, and report the results for two systems (one selected from each dataset due to space limitations) separately. Similar to mediation analysis, the moderation effect of LOC on the DIT metric is excluded as it was insignificant (see section 4.2).

Table 6 presents the moderation effect results for Apache Lucene (v. 2.4) and Apache POI from the DAMB and JURE datasets, respectively. The moderation effect of class size on the relationship between individual OO metrics and the number of faults is significant when the interaction effect ($\rho$) is significant ($\rho < .05$). From Table 6, our moderation analysis reveals that the interaction effects for Lucene 2.4 were significant for all metrics. We conclude that there exist significant moderation effects on all metrics. Similarly, for POI, we observed significant moderation effects of LOC on the individual OO metrics excluding the RFC metric ($\rho = .08$). Furthermore, we investigate the conditional effect to find out when moderation was really significant, that is when the effect of class size (LOC) is low, moderate or high. Conditional effects are significant when $\rho < .05$ and zero does not lie between the BootLL and BootUL confidence intervals.

From the Table, a $\rho < .05$ with positive effect values indicates a significant and positive association between the metric and number of faults (i.e., an increase in the OO metric leads to an increase in the number of faults) whereas a negative effect value indicates that the relationship between the metric and faults is negative (i.e., when the value of the OO metric increases, the number of faults decreases). For an insignificant p-value, it indicates that there is no significant relationship between the OO metric and number of faults. There were instances when the overall interaction effect was significant but none of the conditional effects were significant. An example is WMC and LCOM metrics under the Lucene 2.4 and POI projects respectively. We observe very strong moderation effects at all levels of LOC on the WMC metric for the POI project.

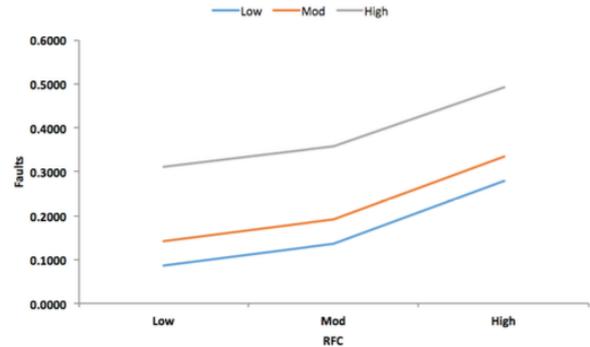

Figure 7: Moderation effect of size on the relationship between RFC and faults in Eclipse Mylyn (DAMB dataset)

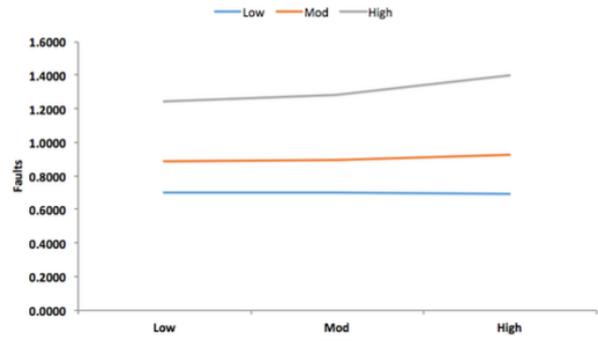

Figure 8: Moderation effect of size on the relationship between WMC and faults in Apache POI (JURE datasets)

We then conducted simple slopes analysis to illustrate the moderation effects of LOC on all OO metrics. Due to space restrictions, we present only the moderation effects on the relationship between RFC and faults for Eclipse Mylyn (DAMB dataset) in Fig. 7, and WMC and faults for Apache POI (JURE dataset) in Fig. 8. We include the graphs from all other system in the supplementary material [24]. As shown in Fig. 7, for all levels of LOC, as the RFC value increases, the number of faults also increases. The moderation effect was significantly strong when the LOC was high. This suggests that the RFC metric has a high probability of explaining faultiness in large classes (LOC is high). Similarly, the results for WMC in Fig. 8 show that with an increase in size (high LOC), the moderation effect becomes strong. However, the lines for moderate and low levels of LOC are flat which indict that there is no relationship between WMC and number of faults (that is, as WMC increases, the number of faults remains unchanged).

The moderation analysis results for all systems in both datasets are presented in Table 7. We only provide the summarized results, which show whether size has a moderation effect on each single metric. Five systems in total show significant mediation effects for all metrics (i.e., Mylyn, Lucene (both 2.2 and 2.4), Ant and Log4J). Alternatively, three systems (i.e., Ivy, Prop and Tomcat) show no moderation effect of class size in any of the examined metrics (Prop also show no mediation effect). The moderation effect of size on WMC and CBO metrics was significant in 11 systems across both datasets, while the effect on LCOM was significant in 10 systems.

Table 6: Moderation effect of size on the relationship between individual OO metrics and Faults for Lucene v. 2.4 (DAMB Dataset) and POI (JURE Dataset)

| Metric | Interaction (moderation effect) | LOC | Lucene 2.4 Effect | p | BootLLCI | BootULCI | Interaction (moderation effect) | POI Effect | p | BootLLCI | BootULCI |
|---|---|---|---|---|---|---|---|---|---|---|---|
| **RFC** | 0.00 (✓) | Low | 0.0035 | 0.00 | 0.0022 | 0.0049 | 0.08 (X) | 0.0329 | 0.00 | 0.0259 | 0.0398 |
|  |  | Med | 0.0036 | 0.00 | 0.0023 | 0.0049 |  | 0.0331 | 0.00 | 0.0263 | 0.0399 |
|  |  | High | 0.0040 | 0.00 | 0.0027 | 0.0053 |  | 0.0339 | 0.00 | 0.0277 | 0.0400 |
| **WMC** | 0.00 (✓) | Low | -0.0019 | 0.17 | -0.0048 | 0.0009 | 0.00 (✓) | 0.0302 | 0.00 | 0.0136 | 0.0470 |
|  |  | Med | -0.0019 | 0.19 | -0.0047 | 0.0009 |  | 0.0323 | 0.00 | 0.0161 | 0.0484 |
|  |  | High | -0.0016 | 0.26 | -0.0044 | 0.0012 |  | 0.0394 | 0.00 | 0.0248 | 0.0540 |
| **CBO** | 0.00 (✓) | Low | 0.0011 | 0.44 | -0.0017 | 0.0038 | 0.00 (✓) | 0.0048 | 0.34 | -0.0050 | 0.0146 |
|  |  | Med | 0.0021 | 0.13 | -0.0006 | 0.0048 |  | 0.0076 | 0.12 | -0.0019 | 0.0170 |
|  |  | High | 0.0061 | 0.00 | 0.0036 | 0.0087 |  | 0.0177 | 0.00 | 0.0093 | 0.0261 |
| **LCOM** | 0.02 (✓) | Low | 0.0008 | 0.00 | 0.0005 | 0.0011 | 0.00 (✓) | -0.0002 | 0.59 | -0.0011 | 0.0062 |
|  |  | Med | 0.0008 | 0.00 | 0.0005 | 0.0011 |  | -0.0002 | 0.65 | -0.0010 | 0.0065 |
|  |  | High | 0.0008 | 0.00 | 0.0005 | 0.0011 |  | -0.0000 | 0.95 | -0.0008 | 0.0072 |
| **Fan-in** | 0.00 (✓) | Low | -0.0004 | 0.77 | -0.0034 | 0.0025 | 0.01 (✓) | 0.0016 | 0.79 | -0.0101 | 0.0134 |
|  |  | Med | 0.0009 | 0.52 | -0.0019 | 0.0038 |  | 0.0038 | 0.49 | -0.0071 | 0.0147 |
|  |  | High | 0.0064 | 0.00 | 0.0037 | 0.0091 |  | 0.0117 | 0.02 | 0.0021 | 0.0214 |
| **Fan-out** | 0.00 (✓) | Low | -0.0091 | 0.09 | -0.0195 | 0.0013 | 0.00 (✓) | 0.0028 | 0.84 | -0.0235 | 0.0291 |
|  |  | Med | -0.0058 | 0.26 | -0.0160 | 0.0044 |  | 0.0080 | 0.54 | -0.0173 | 0.0333 |
|  |  | High | 0.0072 | 0.14 | -0.0023 | 0.0167 |  | 0.0268 | 0.02 | 0.0483 | 0.0488 |

Table 7: Potential moderation effect of class size in all individual systems from both DAMB and JURE datasets

|  | DAMB dataset | | | | | JURE dataset | | | | | | | | | | | |
|---|---|---|---|---|---|---|---|---|---|---|---|---|---|---|---|---|---|
|  | JDT | Mylyn | PDE | Equinox | Lucene (2.4) | Ant | Ivy | jEdit | Log4j | Lucene (2.2) | POI | Prop | Synapse | Tomcat | Velocity | Xalan | Xerces |
| RFC | X | X | X | ✓ | ✓ | ✓ | X | X | X | ✓ | X | X | X | ✓ | ✓ | ✓ | X |
| WMC | X | X | X | X | X | ✓ | X | X | X | ✓ | X | X | X | ✓ | ✓ | ✓ |
| CBO | ✓ | ✓ | ✓ | ✓ | ✓ | ✓ | X | X | X | ✓ | ✓ | X | ✓ | ✓ | X | ✓ | ✓ |
| LCOM | X | X | X | X | ✓ | ✓ | ✓ | X | X | ✓ | ✓ | X | X | X | ✓ | ✓ | ✓ |
| Fan-in | ✓ | ✓ | ✓ | ✓ | ✓ | ✓ | X | X | X | X | X | X | ✓ | X | X | ✓ |
| Fan-out | ✓ | ✓ | X | ✓ | ✓ | ✓ | X | X | X | ✓ | X | ✓ | N/A | ✓ | ✓ | ✓ |

By analyzing the moderation effect of class size in both datasets we found some evidence of a significant moderation effect of size for the WMC, CBO, LCOM and Fan-in metrics. Similar to the mediation effect results, this moderation effect is not consistent when we look at individual systems. Therefore, we contend that the evidence provided regarding the moderation effect of size is not conclusive in regard to these two datasets. Class size seems to moderate the relationship between RFC and faults in 4 out of 5 systems in DAMB dataset, but this was not the case in the JURE dataset (only 3 out of 12 systems). For Fan-in, the moderation effect was also significant in 4 systems in the DAMB dataset, but was evident in only 5 of the 12 systems in the JURE dataset. In returning to our research hypotheses (H1), we cannot reject the null hypothesis that class size has no significant moderation effect on the relationship between the number of faults and the following OO metrics: RFC and Fan-out. For WMC, CBO, LCOM and Fan-in metrics, the evidence available of this effect is not consistent across both datasets. Again, and similar to the mediation effect, we are unable to confirm if the moderation effect will always be present.

### 4.4 Implications of the results

In general, this study provides results contradictory to those of El Emam et al. [6] and Zhou et al. [25]. Based on our in-depth investigation of the mediation and moderation effect of size we cannot confirm if size is actually the reason for the relationship between OO metrics and faults. The main implication of these findings is that, rather than recommending that size should always be controlled when building prediction models, we recommend that one should consider controlling for size only when significant evidence of mediation or moderation effects is observed, and that should be tested for each system individually. Size remains an important metric for predicting faults (and other quality artefacts), but without proper investigation of the mediation and moderation effect of size for individual systems, size should not be omitted or controlled as a rule. Although we have studied the effect on number of faults rather than fault-proneness [6, 25], our results provide a complementary perspective on the effect of class size, indicating that there is no strong evidence of a significant indirect effect of size in fault prediction models. We note that the size effect depends on the systems examined. This is also the case for the moderation effect of size, where the findings regarding such effects are inconsistent across the systems studied. However, there are some indications that size might have a significant moderation effect for some specific metrics. It does appear that the selection of datasets (the systems examined) plays a significant role in determining whether size has a mediation or moderation effect on the association between OO metrics and the number of faults.

We therefore recommend that if evidence of a significant mediation or moderation effect of size is found (which, as this paper advocates, should be examined using robust statistical methods), then one should consider controlling for size before looking at the prediction power of other OO metrics. In such cases, failing to control for size might result in models that are unstable and misleading in interpreting the prediction power of OO metrics.

Still, this does not mean that we should completely ignore (or even overestimate) the role of class size in such models. One may argue that size metrics (e.g., LOC) can be used solely to predict the likelihood of faults given that size is correlated with the number of faults. However, we believe that predicting faults using size alone can also be misleading. Other metrics should also be incorporated when such models are built. The mediation and moderation analysis can reveal what is the impact of size on other metrics. If such an impact is found to be significant (i.e., significant mediation/moderation effect) then a method to control for size should be used (see, for example, [9, 25]). However, class size can be used, with caution, as an initial indicator of the presence of faults in a class. LOC, for example, could be used as a good and quick predictor of faults, as it is generally easier to collect compared to other complexity metrics [10].

## 5. CONCLUSION

Previous studies on faults-proneness have shown that class size has an indirect effect on the relationships between OO metrics and fault-proneness. However, no studies have attempted to study the same phenomena in fault prediction models. Our study reports on the analysis of the mediation and moderation effect of class size on the relationships between OO metrics and the number of faults. When gauging indirect effects, bootstrapping-based techniques have several advantages over other methods, including the casual- steps and Sobel tests. By applying bootstrapping mediation we have shown that there is no conclusive evidence that size has a significant mediation effect on the relationships between most OO metrics and the number of faults, in contrast to previous findings in fault-proneness models. We also found inconsistent evidence that size has a moderation effect (i.e., affects the magnitude of the relationships between OO metrics and the number of faults), with a significant moderation effect found for WMC and CBO metrics.

Unlike previous studies in fault-proneness, our results do not support the assertion that the size effect always exists in continuous fault prediction models. In fact, our results suggest a contrary conclusion - that class size has no significant mediation or moderation effect on the relationships between most OO metrics and the number of faults.

We contend that bootstrapping-based techniques provide more powerful means for analyzing mediator and moderator variables and we therefore recommend that empirical software engineering researchers employ such techniques when studying mediation and moderation effects.

## REFERENCES

[1] Reuben M Baron and David A Kenny. 1986. The Moderator–Mediator Variable Distinction in Social Psychological Research: Conceptual, Strategic, and Statistical Considerations. Journal of personality and social psychology 51, 6 (1986), 1173.

[2] Victor R Basili, Lionel C. Briand, and Walcélio L Melo. 1996. a Validation of Object-Oriented Design Metrics As Quality Indicators. IEEE Transactions on Software Engineering 22, 10 (1996), 751–761.

[3] KwabenaEboBennin,JackyKeung,PassakornPhannachitta,AkitoMonden,and Solomon Mensah. 2018. MAHAKIL: Diversity based Oversampling Approach to Alleviate the Class Imbalance Issue in Software Defect Prediction. IEEE Transactions on Software Engineering 44, 6 (2018).

[4] Kenneth Bollen and Robert Stinet. 1990. Direct and Indirect Effects: Classical and Bootstrap Estimates of Variability. Sociological Methodology 20 (1990), 115–140.

[5] Marco D'Ambros, Michele Lanza, and Romain Robbes. 2012. Evaluating Defect Prediction Approaches: A Benchmark and an Extensive Comparison. Empirical Software Engineering 17, 4-5 (2012), 531–577.

[6] Kalhed El Emam, Saïda Benlarbi, Nishith Goel, and Shesh N. Rai. 2001. The Confounding Effect of Class Size on the Validity of Object-Oriented Metrics. IEEE Transactions on Software Engineering 27, 7 (2001), 630–650.

[7] William M Evanco. 2003. Comments on "The Confounding Effect of Class Size on the Validity of Object-Oriented Metrics". IEEE Transactions on Software Engineering 29, 7 (2003), 670–672.

[8] Matthew S Fritz and David P Mackinnon. 2007. Required Sample Size to Detect the Mediated Effect. Psychological Science 18, 3 (2007), 233–239.

[9] Yossi Gil and Gal Lalouche. 2017. On the Correlation Between Size and Metric Validity. Empirical Software Engineering 22, 5 (2017), 2585–2611.

[10] Tibor Gyimothy, Rudolf Ferenc, and Istvan Siket. 2005. Empirical Validation of Object-Oriented Metrics on Open Source Software for Fault Prediction. IEEE Transactions on Software Engineering 31, 10 (2005), 897–910.

[11] TracyHall,SarahBeecham,DavidBowes,DavidGray,andSteveCounsell.2011. A Systematic Review of Fault Prediction Performance in Software Engineering. IEEE Transactions on Software Engineering 38, 6 (2011), 1276–1304.

[12] AndrewF.Hayes.2009.BeyondBaronandKenny:StatisticalMediation Analysis in the New Millennium. Communication Monographs 76, 4 (2009), 408–420.

[13] Andrew F Hayes. 2013. Introduction to Mediation, Moderation, and Conditional Process Analysis: A Regression-based Approach. Guilford Press.

[14] AndrewF.HayesandKristopherJ.Preacher.2014.StatisticalMediation Analysis with a Multicategorical Independent Variable. Brit. J. Math. Statist. Psych. 67, 3 (2014), 451–470.

[15] Zhimin He, Fengdi Shu, Ye Yang, Mingshu Li, and Qing Wang. 2012. An investigation on the feasibility of cross-project defect prediction. Automated Software Engineering 19, 2 (2012), 167–199.

[16] MarianJureczkoandLechMadeyski.2010.TowardsIdentifyingSoftwareProject Clusters with Regard to Defect Prediction. In Proceedings of the 6th International Conference on Predictive Models in Software Engineering. ACM.

[17] Barbara Kitchenham, Lech Madeyski, David Budgen, Jacky Keung, Pearl Brereton, Stuart Charters, Shirley Gibbs, and Amnart Pohthong. 2016. Robust Statistical Methods for Empirical Software Engineering. Empirical Software Engineering 21, 1 (2016), 212–259.

[18] David P MacKinnon, Chondra M Lockwood, Jeanne M Hoffman, Stephen G West, and Virgil Sheets. 2002. A Comparison of Methods to Test Mediation and Other Intervening Variable Effects. Psychological Methods 7, 1 (2002), 83–104.

[19] David P. MacKinnon, Chondra M. Lockwood, and Jason Williams. 2004. Comparison of Approaches in Estimating Interaction and Quadratic Effects of Latent Variables. Multivariate Behavioral

Research 39, 1 (2004), 37–67.

[20] Kristopher J Preacher and Andrew F Hayes. 2008. Asymptotic and Resampling Strategies for Assessing and Comparing Indirect Effects in Multiple Mediator Models. Behavior Research Methods 40, 3 (2008), 879–891.

[21] Kristopher J Preacher, Derek D Rucker, and Andrew F Hayes. 2007. Addressing Moderated Mediation Hypotheses: Theory, Methods, and Prescriptions. Multi-variate Behavioral Research 42, 1 (2007), 185–227.

[22] Kristopher J Preacher and James P Selig. 2012. Advantages of Monte Carlo Confidence Intervals for Indirect Effects. Communication Methods and Measures 6, 2 (2012), 77–98.

[23] Martin Shepperd, David Bowes, and Tracy Hall. 2014. Researcher Bias: The Use of Machine Learning in Software Defect Prediction. IEEE Transactions on Software Engineering 40, 6 (2014), 603–616.

[24] Amjed Tahir, Kwabena E. Bennin, Stephen G. MacDonell, and Stephen Marsland. 2018. Supplementary material for "Revisiting the size effect in software fault prediction models". https://zenodo.org/record/1317919#.W1JlANgza9b.

[25] Yuming Zhou, Baowen Xu, Hareton Leung, and L I N Chen. 2014. An In-Depth Study of the Potentially Confounding Effect of Class Size in Fault Prediction. ACM Transactions on Software Engineering and Methodology 23, 1 (2014).

[26] Thomas Zimmermann, Rahul Premraj, and Andreas Zeller. 2007. Predicting Defects for Eclipse. In Proceedings of the 3rd International workshop on Predictor Models in Software Engineering. 76.